\documentclass[twocolumn,showpacs,amssymb,aps,nofootinbib,floatfix]{revtex4-1}
\usepackage{epsfig}
\newcommand{\be}{\begin{eqnarray}}
\newcommand{\ee}{\end{eqnarray}}
\newcommand{\ave}[1]{\left\langle #1 \right\rangle}
\newcommand{\mod}[1]{\left| #1 \right|}

\newcommand{\order}[1]{ \mathcal{O} \left( #1 \right) }

\begin{document} \hbadness=10000
\topmargin -0.8cm\oddsidemargin = -0.7cm\evensidemargin = -0.7cm
\title{Rapidity scaling of multiplicity and flow in weakly and strongly interacting systems}
\author{Giorgio Torrieri}
\affiliation{FIAS,
  J.W. Goethe Universit\"at, Frankfurt A.M., Germany  \\
torrieri@fias.uni-frankfurt.de}
\date{November 4, 2010}  

\begin{abstract}

We examine the ``naturalness'' of the scaling of multiplicity and 
elliptic flow $v_2$ with rapidity in weakly and strongly interacting 
systems.  We argue that multiplicity scaling is relatively 
straight-forward to incorporate in existing ansatze, and that this scaling is insensitive to 
the transport properties of the system.
On the other hand, we argue that the observed scaling of elliptic flow 
is problematic to describe within a hydrodynamic model (the Knudsen number $K \ll 1$), but 
arises more naturally within weakly interacting systems (where the Knudsen 
number $\sim 1$).  We conclude by an overview of ways proposed to make 
weakly interacting systems compatible with the absolute value of elliptic 
flow, and 
by indicating experimental probes which could clarify these issues.
\end{abstract}
\maketitle
\section{introduction}
The azimuthal anisotropy of mean particle momentum (parametrized by it's
second Fourier component $v_2$), thought of as originating from the
azimuthal anisotropy in collective flow (``elliptic flow''), has long been regarded as an
important observable in heavy ion collisions.
The main reasons for this is that elliptic flow has long been thought
to be ``self-quenching'' \cite{v2orig,v2orig2}: The azimuthal pressure gradient
extinguishes itself soon after the start of the hydrodynamic evolution, so the final
$v_2$ is insensitive to later stages of the fireball evolution and
therefore allows us to probe the hottest, best thermalized, and
possibly deconfined phase.

In addition, as has been shown in \cite{teaney}, elliptic flow
is highly sensitive to viscosity.  The presence of even a small but
non-negligible viscosity, therefore, can in principle be detected by a
careful analysis of $v_2$ data.

Indeed, one of the most widely cited (in both the academic and popular
press) news coming out of the heavy ion community concerns the discovery, at the relativistic heavy ion collider ( RHIC ), of a ``perfect fluid'',
also sometimes referred to as ``sQGP'' (strongly coupled Quark Gluon Plasma)
\cite{whitebrahms,whitephobos,whitestar,whitephenix,sqgpmiklos,sqgpshuryak}.
The evidence for this claim comes from the successful
modeling of RHIC $v_2$ by boost-invariant hydrodynamics
\cite{tauv2,shuryak,romatschke,huovinen,chojnacki,hirano}.   The scaling of $v_2$ according 
to the 
number of constituent quarks further suggests that the flow we are seeing is partonic in origin \cite{npart1,npart2,npart3,npart4,taranenko}.

Going further in our understanding is hampered by the large number of ``free'' (or, to be more exact, poorly understood from first principles) parameters within the hydrodynamic model: While the equation of state is thought to be understood from lattice simulations, the behavior of shear and bulk viscosity is quantitatively not known around $T_c$, where it is expected the temperature dependence could be non-trivial \cite{denicol,bozvisc,bulkvisc1,bulkvisc2}.   The same goes for the large number of second-order transport coefficients.   While we have some understanding of the initial transverse density of the system (its dependence on the transverse coordinate is thought to be either a ``Glauber'' superposition of p-p collisions \cite{hsong} or a partonic semi-classical ``color glass'' \cite{cgc,kln}), we do not as yet have control over the degree of transparency of the system, the amount of transverse flow created before thermalization (thought to be necessary to make the data agree with particle interferometry measurements \cite{pratt}), or of the interplay between the ``medium'' and the surrounding ``corona'' of peripheral p-p collisions \cite{corecorona1,corecorona2}.   A model incorporating ``all physics'', therefore, is expected to have a lot of correlated parameters which will be highly non-trivial to disentangle.

A tool with the potential of overcoming these difficulties is scaling naturalness.  
Experiments have collected an extraordinary amount of flow data, encompassing a wide range of Energy ($\sqrt{s}$),centrality (parametrized by number of participants $N_{part}$), system size (mass number $A$ of the nuclei), rapidity $y$ and pseudorapidity $\eta$, particle species and transverse momentum ($p_T$).   

The experimental data collected allows us to ``scan'' observables dependence on variables relevant to the theory, and to see if the observable change when the same variable is obtained in different ways (for example, flow at mid-rapidity of a lower $\sqrt{s}$ collision compared with flow at the fragmentation region of a higher $\sqrt{s}$ collision,at the same multiplicity density $dN/dy$).

This is an important test, because many theories have only one such relevant scaling variable (RSV) expected to drive the observable's  value. Conversely, many theories have several RSVs, which combine to produce the observable's value.

In the first case, one expects the observable to depend on {\em only} one scaling parameter in a straight-forward manner, no matter how the scaling parameter was arrived at (eg, flow should depend on $dN/dy$ only, no matter weather $dN/dy$ corresponds to mid-rapidity at low $\sqrt{s}$ or high rapidity at high $\sqrt{s}$).
In the second case, one expects the scaling to fail (flows for the same $dN/dy$ but different rapidity/$\sqrt{s}$ will be different), because the RSVs required to produce the scanned parameter will generally be different as one scans in energy,system size, centrality etc.
If this expectation is not fulfilled, complicating the model (e.g., adding more transport coefficients to the hydrodynamic model) is {\em not} likely to create a scaling where there was none, because the extra features generally also add RSVs \footnote{As an example of how this works, a very popular, and physically reasonable refinement to models describing heavy ion collisions is the ``core-corona model'', assuming that the bulk of the system, evolving collectively, is surrounded by a peripheral corona which is an incoherent superposition of $p-p$ collisions \cite{corecorona1,corecorona2}.  While one can implement this in many different ways, from the assumptions of the model it follows that more central collisions have more ``core'' and less ``corona'', while collisions with a different $\sqrt{s}$ have a similar percentage of ``core'' and ``corona'' but different dynamics for ``core'' {\em and} corona.   Hence, if one compares data at different centralities and multiplicities but same $dN/dy$, one should not expect any kind of scaling of observables with $dN/dy$ in models incorporating core and corona dynamics.  Conversely, if such a scaling is found, it is an indication that ``core''/''corona'' separation does not play a big role in the physics}.

If small dimensionless quantities (SDQs) $\alpha_i$ can be constructed from some of the RSVs, the above statement is mathematically equivalent to Taylor-expanding the flow observable around the SDQs \cite{scaling}
\begin{equation}
\label{expdef}
\ave{observable} \sim A_0 +A_1 \alpha_1 + B_2 \alpha_2  + \order{\alpha_1 \alpha_2}+ ...
\end{equation}
where $A_i$,to leading order,are independent of $\alpha_i$.

In case of hydrodynamics,the parameters relevant for the dynamics are expected to be the initial temperature and chemical potential (setting the speed of sound $c_s$, the free path $l_{mfp}$),the initial system size, and the intrinsic scales 
of the theory.  In practice, the chemical potential is thought to be irrelevant for higher energy collisions, and the intrinsic scales of QCD combine to produce one thermal scale $T_c$, thought to coincide with a dip in the sound velocity and a change in the viscosity over entropy density ratio. $\eta/s$ (the dip and the change would be sharp for a phase transition, or smooth for a cross-over).   These relevant parameters can be accommodated in SDQs,but
\begin{itemize}
\item  more than one SDQ is necessary: At least Knudsen number $K$,eccentricity $\epsilon$,speed of sound $c_s$
\item A dimensionful theoretical scale ($T_c$) appears in the relevant SDQs: $c_s$ and $K$ depend on $T$,with a scale given by $T_c$
\item some SDQs are purely intensive (the speed of sound) while others (the Knudsen number) mix intensive and extensive variables
\end{itemize}
the last two points mean scanning is possible, given a wide enough space in $\sqrt{s}$, centrality and rapidity.  The first point means
flow observables are {\em not } automatically expected to scale easily with the multiplicity, since several combinations of relevant variables (higher initial temperature vs larger system) can underlie the same multiplicity, yet produce different dynamics.

These considerations can be promoted to a test of the hydrodynamic model, as several
``simple'' patterns have been found, centered around the flow at different energies, system sizes and rapidities scaling with a few global dimensionless parameters (such as eccentricity and multiplicity density). 
As we will show, in fact, such simple dependence is generally incompatible with an expansion such as Eq. \ref{expdef} in the context of hydrodynamics with a QCD equation of state, but is more compatible with a {\em weakly} interacting system,where the Knudsen number is of order unity (each degree of freedom interacts only once in the system's lifetime).

We shall illustrate these points with one observable not yet fully examined within viscous hydrodynamic models: the $v_2$ dependence on particle rapidity $y=\tanh^{-1} p_z/E$ or pseudorapidity $\eta= \tanh^{-1}p_z/\mod{p}$,where $p_z$ is the momentum in the beam direction and, $y \simeq \eta$ for $p_z \gg p_T,m$, true for most particles away from mid-rapidity\footnote{Unfortunately, as in the common notation, $\eta$ can mean both viscosity and pseudorapidity, $s$ can mean both the square of the center of mass energy and the entropy density, while $S$ is the nuclear overlap area.  The reader unfamiliar with the field should watch for the context in which these letters are used to understand their meaning  }.

The reason for the comparative lack of interest for $v_2$'s dependence of rapidity is both numerical and physical: Numerically, introducing a rapidity dependence in the system makes the problem much harder to solve.
Physically, it was thought that at {\em mid-} rapidity Boost-invariant initial condition \cite{bjorken} is a good description.   Since this is also the 
denser 
region, where the ``hot'' QGP phase will be most prominent, concentrating on this region and relying on Boost-invariance to shield it from fragmentation effects in the projectile and target rapidity is a priori a sensible approach.

Experimental data, however, seems to show at best a very narrow central plateau for both the multiplicity distribution \cite{dndybrahms,busza} and elliptic flow \cite{busza,rapbrahms}.  While data for rapidity dependence close to mid-rapidity is somewhat ambiguous (the PHOBOS collaboration reports a triangular shape \cite{whitephobos}, while the STAR and PHENIX collaborations suggest a trapezium following the multiplicity distribution\cite{whitestar,whitephenix}), it is clear that $v_2$ starts decreasing \cite{whitephobos,whitebrahms} far from what is naively thought of as the unthermalized fragmentation region.    

Even more tantalizing are the systematics in both energy and rapidity of these observables:   It appears that limiting fragmentation \cite{limfra} holds as well for A-A collisions \cite{unifrag}  as for more elementary systems.  Multiplicity at mid-rapidity also seems to be ``aware'' of limiting fragmentation, since it follows, with no ``sudden'' transitions, the trend \cite{busza}
\begin{equation}
\label{dndysyst}
\left. \frac{dN}{dy} \right|_{y=0} \sim N_{part}\ln \sqrt{s}
\end{equation}
where $N_{part}$ is the number of participants and $\sqrt{s}$ is the center of mass energy.

Even more puzzling, $v_2$ for A-A systems follows these trends as well, both w.r.t. to scaling with pseudo-rapidity \cite{busza} and at mid-rapidity \cite{scaling}.

Fig. \ref{figeta} summarizes these findings.
It is the purpose of the current work to assess these experimental observations in the light of current models used to understand RHIC data.
We should underline that we are analyzing the {\em scaling} rather than the {\em absolute values} of these observables.   Detailed studies of the latter have been made at top RHIC energies \cite{v2rap1,v2rap2,v2rap3,v2rap4,v2rap5}, and it was found that models combining ideal hydrodynamics and hadronic cascades reproduce multiplicity pretty well, and do a reasonable through not perfect job of reproducing $v_2$. Scaling of these observables, within a theory containing intensive dimensionful parameters, has not however been quantitatively tested, and could hold the key to see whether data-model disagreements seen in \cite{v2rap1,v2rap2,v2rap3,v2rap4,v2rap5} are compatible with experimental systematic errors, can be fixed with minor model refinements (coronae, rescattering, a more realistic initial state,...) or whether they indicate a more profound misunderstanding of the system under consideration.

Section \ref{secdn} aims to demonstrate that scaling of $dN/dy$ (both in the limiting fragmentation region) and at mid-rapidity, is generally compatible with expectations from high-energy QCD, and that it is rather insensitive to the phase of the system at thermalization.  The reader should {\em not} treat this section as a demonstration of a rigorous model, but as an argument that a rigorous model in agreement with what we know from high energy QCD {\em can} be developed, and as an outline of the requirements such a model needs to satisfy to describe the scaling.

The main section of this work, Section \ref{secv2}, demonstrates that the situation is rather different for the limiting fragmentation of elliptic flow, since, unlike $dN/dy$, elliptic flow {\em should} be sensitive to the phase of the system at thermalization, and that therefore the observed fragmentation of $v_2$ is not natural within hydrodynamics

Section \ref{secweak} argues that, in fact, such scaling arises more naturally in a weakly interacting theory, and section \ref{secexp} outlines experimental probes capable of clarifying the situation and closes with concluding remarks.

\section{ \label{secdn}Scaling of $dN/dy$}
\subsection{Limiting fragmentation \label{subsec1}}

 
\begin{figure*}[t]
\epsfig{width=19cm,figure=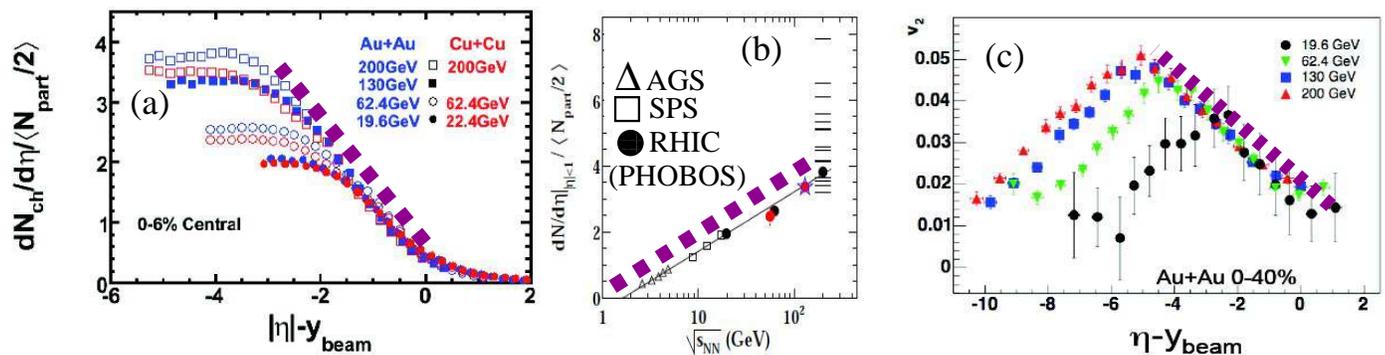}
\caption{\label{figeta} (color online) The experimentally observed multiplicity rapidity distribution (a), the multiplicity at midrapidity as a function of $\sqrt{s}$ (b), and the rapidity dependence of $v_2$ (c).
(a) is usually known as limiting fragmentation and (c) is referred to as limiting fragmentation of elliptic flow.
  All experimental plots from \cite{busza}, with the dashed line indicating the scaling trend}
\end{figure*}

\begin{figure*}[t]
\epsfig{width=19cm,figure=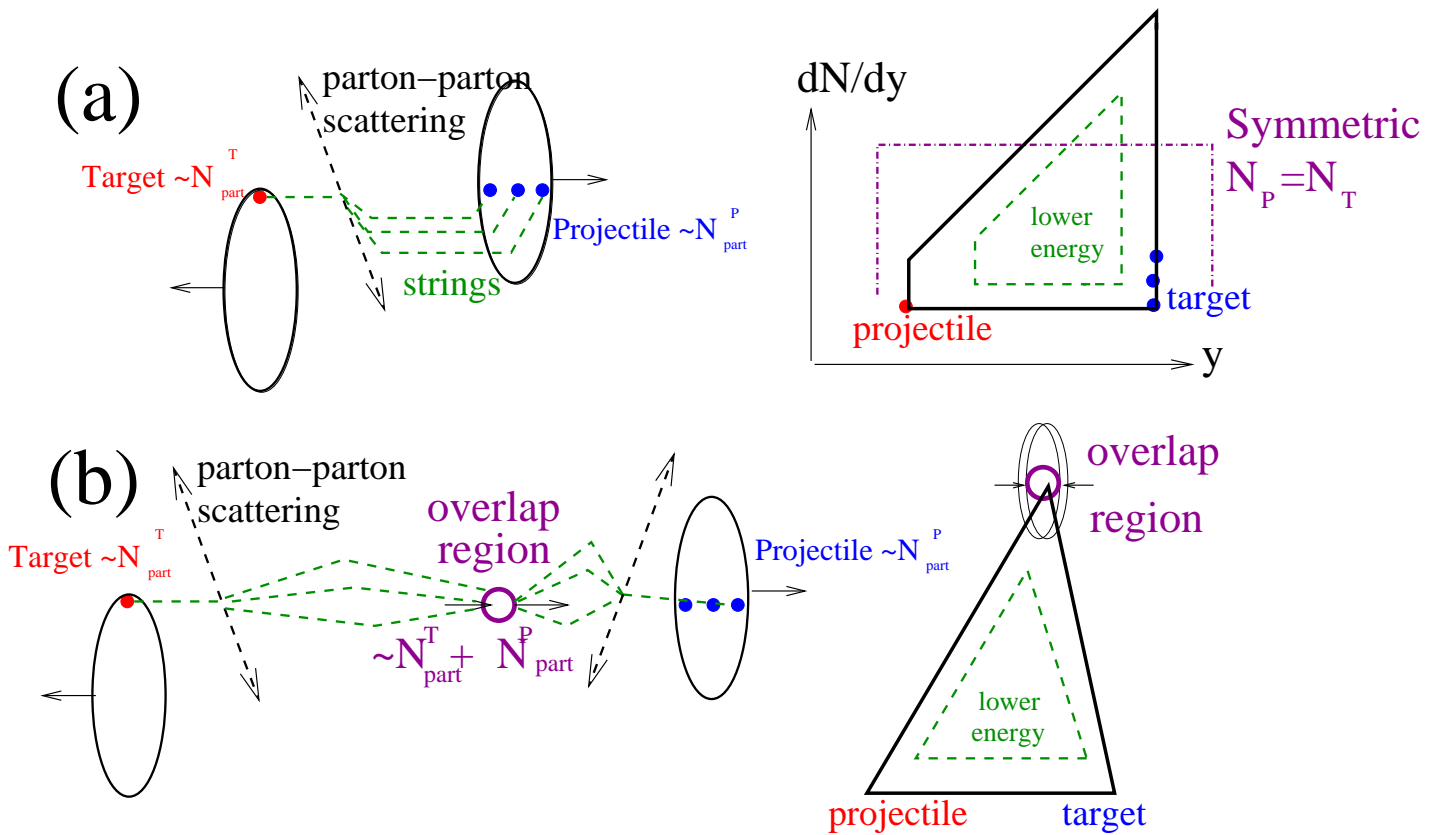}
\caption{\label{figbgk} (color online) The ``usual'' BGK picture discussed in section \ref{subsec1} (a) 
and the modified BGK picture discussed in section \ref{subsec2} (b).  In the usual picture, the partons 
of the projectile and target (denoted by red and blue filled circles) 
act as origin points of color flux tubes (denoted as green dashed 
lines).  Partons from each flux tube are generated in a boost invariant configuration, but, after summing over all flux tubes, parton density is sensitive to the ratio of wounded nucleons in the 
projectile and the target, according to a linear extrapolation.   The 
phenomenological modification in the   (b) can be understood in 
terms of a third dominant source of color flux tubes, located at the spacetime locus of the collision between the projectile and target (denoted by the larger hollow magenta circle)  } 
\end{figure*}

While an exhaustive explanation for limiting fragmentation \cite{limfra} still evades 
us, it is a plausible feature of many models easily justified within the 
QCD paradigm, in particular in the models based around the ``wounded nucleon'' or ``wounded quark'' concept \cite{wounded} (essentially, a superposition of independent collisions each producing elongated objects in rapidity).
Without loss of generality, we shall concentrate on the Brodsky-Gunion-Kuhn (BGK) \cite{bgk,gyuladil} model in its 
more generic form (Fig \ref{figbgk}  (a)), to illustrate why limiting fragmentation is plausible, and 
to suggest why the logarithmic dependence of the multiplicity density at 
mid-rapidity can be plausibly understood as being related to limiting 
fragmentation.
While the BGK model is based on 
a generic scenario in terms of parton-parton collisions, its ansatz is 
easily visualized within the flux tube (``QCD string'') picture.  It should be noted that 
the Color Glass coordinate \cite{cgc,kln} initial state follows this 
ansatz, with the crucial addition of transverse dynamics dictated by 
saturation. The ansatz in \cite{bgk} is however more general, and 
therefore likely to be applicable to regions in $\sqrt{s}$ and 
centrality beyond those where the assumptions behind \cite{cgc} are 
reasonable.

In this picture,  each collision between a projectile and target nucleon produces energetic partons, distributed uniformly in rapidity throughout the 
 forward-traveling projectile and the backward-traveling target.  These partons act as ``string ends'',on average for \\$\rho(y_{min}) \sim N_{part}^T$ strings between the parton and the target, and $\rho(y_{max}) \sim N_{part}^P$ strings between the parton and the projectile ($N_{part}^{T,P}$ are the number of participants from, respectively, the target and the projectile).  Medium partons
are produced through fragmentation of these strings, uniformly in rapidity within each string.
 Partons from fragmenting strings are initially flowing with Boost-invariant flow ($v_z=z/t$ so the spacetime rapidity $$ \tanh^{-1} \frac{z_{dyn}}{t_{dyn}} = y$$  (where $z_{dyn},t_{dyn}$ are the formation spacetime points) , as expected in parton generation from longitudinally expanding flux tubes \cite{werner} and confirmed by measurements such as net baryon number in rapidity \cite{rapbrahms} (showing that high $x_F=p_z/E$ valence partons tend to pass through the collision region with very little change in momentum).  However, parton density $\rho$ is {\em not} invariant with $y$, but rather follows from the sum of independent fragmentations of these flux tubes. 

It is easy to see \cite{bgk,gyuladil} that the resulting multiplicity distribution is 
 a linear extrapolation from target (at $y_{max}$) to projectile (at $y_{min}= - y_{max}$ in the center of mass frame)
\begin{equation}
\label{bgk1}
\rho_{BGK} (y) \sim A - B y
\end{equation}
where 
\begin{equation}
\label{bgk2}
\frac{B}{A} = \frac{N_p^{projectile} - N_p^{target}}{2 y_{max}}
\end{equation}
and $B$ depends weakly on $\sqrt{s}$ ($\frac{dB}{d\sqrt{s}} \leq \frac{1}{\sqrt{s}}$), and scales as the number of participants $N_{part}$.

All we did here is illustrate the features a theory should possess to describe limiting fragmentation in a natural way. 
 While a detailed quantitative  explanation of how these features arise in QCD evades us to this day, it is generally thought they are not unreasonable:  As Fig.
 \ref{figbgk}  (a) shows, it is natural to expect $B$ to simply count Flux Tubes
 (be they the old-fashioned 
electric flux tubes \cite{werner} or the magnetic CGC version \cite{glasma}), and these flux tubes to stretch between the liberated parton and the projectile or target rapidity.  It is therefore not surprising that $B$ is only weakly dependent on $\sqrt{s}$

In the projectile=target limit the BGK picture tends to the boost-invariant limit which, even at RHIC, does not seem to arise in experimental data \cite{busza,rapbrahms}.   Furthermore, in this picture, it is not natural to reproduce the scaling $\rho_{BGK}(y=0) \left[ =A \right] \sim \ln \sqrt{s}$ seen experimentally.
The next subsection discusses possible ways to fix this.
\subsection{$\left. \frac{dN}{dy}\right|_{y=0}$ scaling with $\sqrt{s}$ \label{subsec2}}

The most straight-forward interpretation of the $\left. \frac{dN}{dy} \right|_{y=0} \sim N_{part} \ln \sqrt{s}$ scaling is that the limiting fragmentation region reaches until {\em mid-rapidity} (the initial partonic distribution is a {\em triangle}, and any central plateau is dominated by thermal smearing rather than longitudinal flow).    Since $y_{lim} \sim \ln \sqrt{s}$, assuming that the slopes of $dN/dy$ distributions are constant with respect to $\sqrt{s}$, the logarithmic scaling of the mid-rapidity multiplicity density would automatically follow.

This straight-forward interpretation, however, is  {\em not} trivial to reconcile with long-held beliefs regarding high-energy QCD used to develop ansatze such as BGK.   Limiting fragmentation is only natural in a system with little stopping power, since it implies that, in the frame at rest with respect to the target, the projectile becomes a Lorentz-invariant pancake while the target is extended \cite{limfra}.  That is {\em not} the case in the center of mass frame, where both projectile and target should be equally Lorentz-contracted to a size much smaller than their rest-frame size.

Furthermore, limiting fragmentation follows naturally if distributions created by each parton-parton collision are invariant in rapidity, as in the traditional BGK picture (the linear rapidity dependence in non-central collisions comes from combinatorics of the strings created from the target and the projectile, {\em not} from microscopic dynamics). 

This, in turn, is thought to be based on asymptotic freedom, which suppresses multi-GeV momentum transfers necessary to shift degrees of freedom around the required rapidity range.   RHIC experiments have shattered our understanding of asymptotic freedom applied to multi-particle processes, and stimulated the development of several theoretical scenarios  (strong semi-classical fields \cite{classinst}, semi-classical small-$x$ physics \cite{cgc}, strongly coupled quantum fields tractable by Gauge/string methods \cite{strongc} and so on) where liberated partons interact ``strongly''. 
This raises the possibility that the stopping of color charges deviates from the naive BGK-like value.

It is not immediately clear, however, how large stopping at mid-rapidity could produce limiting fragmentation away from mid-rapidity.
One needs not just to break boost-invariance, but to extend the set-up leading to Eq.  \ref{bgk1},\ref{bgk2} so that the rapidity narrow trapezium is centered around mid-rapidity, and not around the projectile/target local transverse parton density.   

 For instance, Landau hydrodynamics \cite{landau} appears to exhibit approximate limiting fragmentation close to the edges of the rapidity distribution, although corrections, even neglecting transverse flow and using the conformal ideal gas EoS rather than the QCD one, are larger than the deviations seen in the data \cite{wong}. But the limiting fragmentation would not extend to anywhere near mid-rapidity, since Landau hydrodynamics at $y \ll y_{lim}$ becomes indistinguishable from Bjorken hydrodynamics \cite{bjorken} after, typically, a few $T_{initial}^{-1}\sim \order{1}/\sqrt{s}$ \cite{landau}.  Consequently, $dN/dy$ at mid-rapidity in Landau hydrodynamics is not logarithmic wrt $\sqrt{s}$ \cite{landau} unless the initial temperature is unnaturally adjusted to produce the logarithmic dependence.

Similarly, weakly coupled transport models such as \cite{bass2} can produce rapidity distributions without plateaus, but generally break limiting fragmentation.  This is not surprising, since limiting fragmentation is not natural in a model with multiple partonic scatterings.

\begin{figure*}[t]
\epsfig{width=14cm,figure=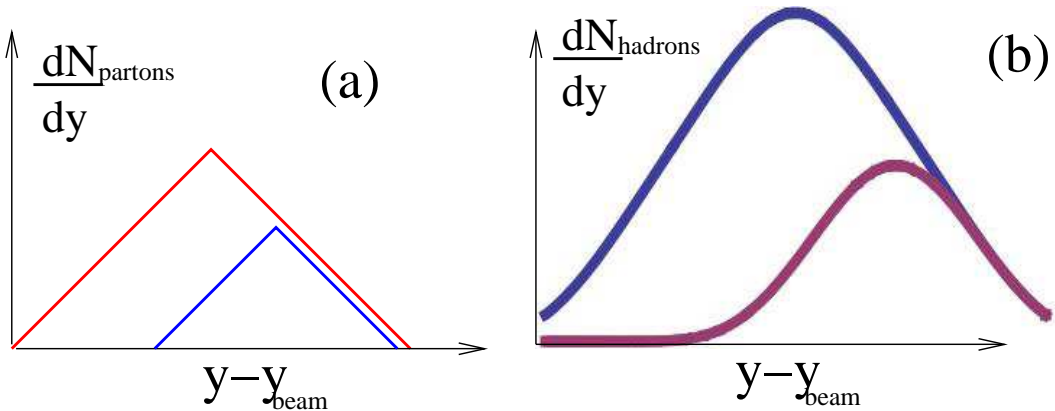}
\caption{\label{figsmear} (color online) (a): The parton density in the modified BGK picture. (b):  The final hadron distribution function, including thermal smearing.}
\end{figure*}

The simplest way to account for the data is to try to tinker with the BGK ansatz 
so that the ``upright trapeziums'' seen in Fig. \ref{figbgk} (a) become triangles described in the first paragraph of this section, shown in Fig. \ref{figbgk} (b).
One sketch of a way in which this could be done within the existing QCD language is to allow for enough ($\sim N_{part}^P+N_{part}^T$) color charge stopping at the center of mass frame \footnote{These slow color charges are sometimes called,and might well be, ``wee partons'' \cite{cgc}. For the discussion here, however, we shall just refer to them as ``color charges'' or ``string ends'', as they might not be partons at all, but some kind of collective excitation in strongly coupled dynamics a la \cite{strongc}.  All we need is that they carry color } to act as a source of color strings along with the ends.    Partons created in the initial scattering (each scattering distributes them uniformly across rapidity, just like in the BGK picture) can then be joined by strings to the projectile, target, {\em and} the central collision region.
If the density of  at the projectile and target ``string ends'' $$\rho(y_{min,max}) \sim N_{part}^{T,P}$$ and in the middle it is $$\rho\left( \frac{y_{min}+y_{max}}{2}\right)\sim N_{part}^{T}+N_{part}^P > \rho \left( y_{min},y_{max} \right)$$ then,following the ansatz used to derive Eqs. \ref{bgk1},\ref{bgk2} of each liberated parton producing a string connecting either the projectile, the target, {\em or} (most likely) the overlap region.  If strings fragment to a boost-invariant distribution,we get
\begin{equation}
\label{modbgk1}
\rho (y) \sim A_{P,T} - B_{P,T} \left| y - y_{cm}  \right|
\end{equation}
where 
\begin{equation}
\label{modbgk2}
y_{cm} \sim \frac{N_p^{projectile} - N_p^{target}}{N_p^{projectile} + N_p^{target}}
\end{equation}
in this ansatz, the initial rapidity distribution becomes not a boost-invariant distribution but a ``triangle'' (Fig. \ref{figsmear}  (a)) ,which thermal smearing can reduce to the observed near-Gaussian (Fig. \ref{figsmear}  (b)).  Furthermore, since $y_{max} \sim \ln \sqrt{s}$, and the slopes from the center of mass to the projectile/target are, respectively, $N_{part}^{P,T}$ (and independent of $\sqrt{s}$). It is easy to see that, for the linear extrapolation to be valid, 
\begin{equation}
\label{modbgk3}
A \sim  \left( N_{part}^P + N_{part}^T \right) \ln \sqrt{s} = N_{part} \ln \sqrt{s} 
\end{equation}
Essentially, the modification of the BGK picture we have made, together with the requirement of limiting fragmentation {\em forces} the scaling seen in Eq. \ref{dndysyst} at mid-rapidity.
Note that in the highly asymmetric limit ($p-A$,or $A-A$ at large transverse radius) this picture tends to the BGK one, since the center of mass also moves with large rapidity .

Fig. \ref{figsmear}  (a) summarize the initial parton distribution at different energies expected from the modified BGK model.
(b) simulates thermal smearing within the Bjorken co-moving frame, through a plot of $f(y-y_L,y_L)$ defined as
\begin{equation}
f(y,y_L) = y_L  \int_{-y_L}^{y_L} d y' \left( 1 - \frac{\left| y' \right|}{y_L} \right) \exp \left[ -\cosh \left( y' -y_L \right)  \right]
\end{equation}
for two representative values of $y_L$.     As can be seen, limiting fragmentation and the mid-rapidity abundance follow the parton distributions very well even after thermal smearing, while the overall rapidity shape looks very similar to the experimental one: The smeared triangle looks like a Gaussian.  A modification of the triangle into a narrow trapezium, suggested by the plateau reported in \cite{whitephobos,whitestar} in both $dN/dy$ and $v_2$ will not modify this conclusion, provided the rapidity width of the tip is not much larger than one unit.

The initial distribution  of string ends in the above scenario  is ``three-pronged'', with peaks in th distribution at $x_f \simeq \pm 1$ (mostly valence quarks, as can be seen from net baryon measurements \cite{rapbrahms}) and a third peak at $x_F =0$ (with width $\ll 1$.  In the standard BGK picture this third peak would not be there).  Strings would then stretch between
either of these peaks, and the newly created partons (evenly distributed in rapidity).

It is not at all clear, and not the aim of this work, to specify which dynamics gives such a distribution of color charges.  The question to answer here is: Why should string ends be created from strongly interacting (not necessarily non-perturbative!) dynamics at the overlap point in configuration space, but only from usual parton-parton scattering and fragmentation elsewhere?   
We only note that, {\em in spacetime rapidity}, the loci of the system where color charges {\em initially} reside are exactly at the ones
corresponding to these peaks: The lightcones of the target and projectile, and the spacetime region where the target and projectile overlap, at the center of mass rapidity.
Thus, the distribution of color charges we conjecture is {\em not} unnatural if strong short-range (cut-off at  a scale $< \Lambda_{QCD}^{-1}$) many-body dynamics at the overlap point is present. 
The gradients of these for high energy collisions are very large $\left( \sim \sqrt{s} N_{part}^{-1/3} \right)$, so, provided the strong many body dynamics is cut off at large distances, color charges between $\left| x_F \right| \ll 1$ and $\left| x_F \right| \simeq 1$ (the partons between the three peaks, which function as other ends of the string) will only be produced by pQCD scatterings, which, up to LHC energies, are subleading to soft particle emission. Strings completely disconnected from all three peaks would therefore be less likely.

Once again, the preceding reasoning should be regarded as a {\em sketch} of a way in 
which both $dN/dy \sim N_{part} \ln \sqrt{s}$ dependence and limiting 
fragmentation arise within the framework of QCD (partons,strings, etc.),or of a way to initialize initial conditions for a dynamical model (hydrodynamics, transport) which respects the scaling seen in experiment.  
It is not a model, since we do not specify which form of 
initial dynamics best satisfies Eq. 
\ref{modbgk1},\ref{modbgk2},\ref{modbgk3}, and because the microscopic dynamics
interpolating between initial partons and later strings is left undetermined. 

But we note that a model which does {\em not} has little chance 
of describing {\em both} limiting fragmentation {\em and} $dN/dy \sim \ln \sqrt{s}$, 
as required to agree with experimental data when a wide range of $\sqrt{s}$ ($\order{10-100}$ GeV) is scanned.

\subsection{$dN/dy$ and the phases of QCD}
It is important to note that here we discussed the {\em initial formation} state of the partons (the start of dynamics $\tau_{dyn}$), not the {\em initial local equilibrium} state of the system (the start of hydrodynamics $\tau_{eq}$, subsequent to string formation, decay of strings into partons, and parton equilibration).  This is why we treated the system as partonic {\em throughout} the rapidity region.

The properties of the medium at $\tau_{eq}$ need to be determined through the initial temperature, obtained the QCD equation of state to link the density to temperature and chemical potential.   Thus, there will be a region in rapidity ($|y|<y_c$) where, at $\tau=\tau_{eq}$, the bulk of the initially equilibrated $\rho$ will be carried by partons, and another region in rapidity where it will be carried by hadrons (put in in a different way, a region where the initial state will be a quark-gluon plasma, and a region where it is a hadron gas.  For some energies the first region dominates, for others the whole system has little or no QGP component).

\begin{figure*}[t]
\epsfig{width=18cm,figure=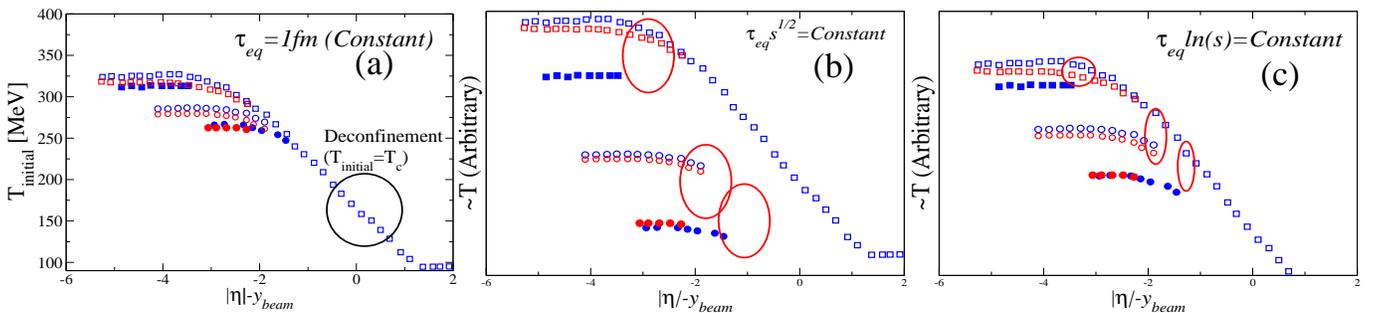}
\caption{\label{dndeta} (color online)  (a): The initial temperature dependence on rapidity for different $\sqrt{s}$, assuming the multiplicity distribution shown in Fig \ref{figeta} (a) (experimental data from \cite{busza}), the Bjorken initial state given by Eq. \ref{teq}, and $\tau_{eq} \sim T_c^{-1}$ .  The symbol shapes and colors follow the coding of Fig. \ref{figeta} (b). 
Panels (b) and (c) show, respectively, the same calculation assuming $\tau_{eq} \sim 1/\sqrt{s}$ and $\tau_{eq} \sim 1/\ln \sqrt{s}$.  The black ellipse in (a) indicates the point where effects due to the phase transition are expected to be most important, while red ellipses in (b),(c) indicate where limiting fragmentation would be expected to break down}
\end{figure*}

To roughly estimate the rapidity at which, on average over the fireball ``slice'', the switch happens, we have used the ideal 3-flavor QCD equation of state to calculate the initial temperature assuming boost-invariant flow ({\em not} full boost-invariance) \cite{bjorken} 
\begin{equation}
\label{teq}
T_{eq} = \left( \frac{dN}{d y}\frac{4 g}{\tau_{eq} N_{part}^{2/3} fm^2} \right)^{1/3} \sim \rho(\tau_{eq})^{1/3}
\end{equation}
where $g=20.8$ is the appropriate degeneracy constant for QCD. The transition point is around where $T_{eq}\sim T_c$,where $T_c \sim 180$ MeV is the critical temperature for the phase transition \cite{jansbook}.
 Fig. \ref{dndeta}  (a) shows the result of this exercise: The critical rapidity for top RHIC energies is $y=3.5-4$, corresponding to a critical $\sqrt{s}$ (where that region is at mid-rapidity) of $\sqrt{s}\sim 3$ GeV (or 18 GeV fixed target, intriguingly close to the ``kink,horn'' and ``step'' observations \cite{horn}).    It should of course be remembered that this is a {\em very} rough approximation, but not an unreasonable one considering that the ansatze described in the previous subsection mean that  the pre-equilibrium medium (at $\sim \tau_{dyn}$) is partonic {\em throughout} the rapidity range.     

The estimate shown in Fig. \ref{dndeta} (a) was made assuming $\tau_{eq} \sim T_c^{-1} \sim 1 fm$, constant w.r.t. $\sqrt{s}$.  This is {\em not} what is usually assumed: By the uncertainty principle, $\tau_{eq}$ should scale as $1/\sqrt{s}$, or at least $1/\ln \sqrt{s}$ if saturation effects dominate.
It should however be noted that any change of $\tau_{eq}$ in energy and system size would generally require an unnatural fine-tuning to reproduce the limiting fragmentation, as panels (b),(c) of the figure show.
As Fig. \ref{dndeta} shows, the ``transition'' region is well within experimental acceptance, and hence any physical effect of it should be evident from a rapidity scan.

To understand what this effect might be, we
note that both nearly free streaming evolution and close-to-ideal liquid expansion conserve entropy.  Therefore, the final $dN/dy$ is relatively insensitive to whether the system was deconfined or not at equilibration.
In other words, if at a given $y$ and $\sqrt{s}$ there is a critical rapidity $y_c$ where a ``perfect liquid'' quark-gluon plasma changes into a ``lousy liquid'' hadron gas, we would not see ``anything special'' when measuring the $dN/dy$ at $y_c$, as entropy conservation would ensure final hadronic multiplicities would mirror initial partonic ones both in a long-lived sQGP ($K \ll 1$) and a short-lived collection of nearly free-steaming hadrons ($K\sim 1$).  Flow observables, and in particular $v_2$, should however be very sensitive to such changes.  The next section will demonstrate these points explicitly.
\section{Naturalness of scaling of elliptic flow in hydrodynamics \label{secv2}}

\begin{figure*}[t]
\epsfig{width=18cm,figure=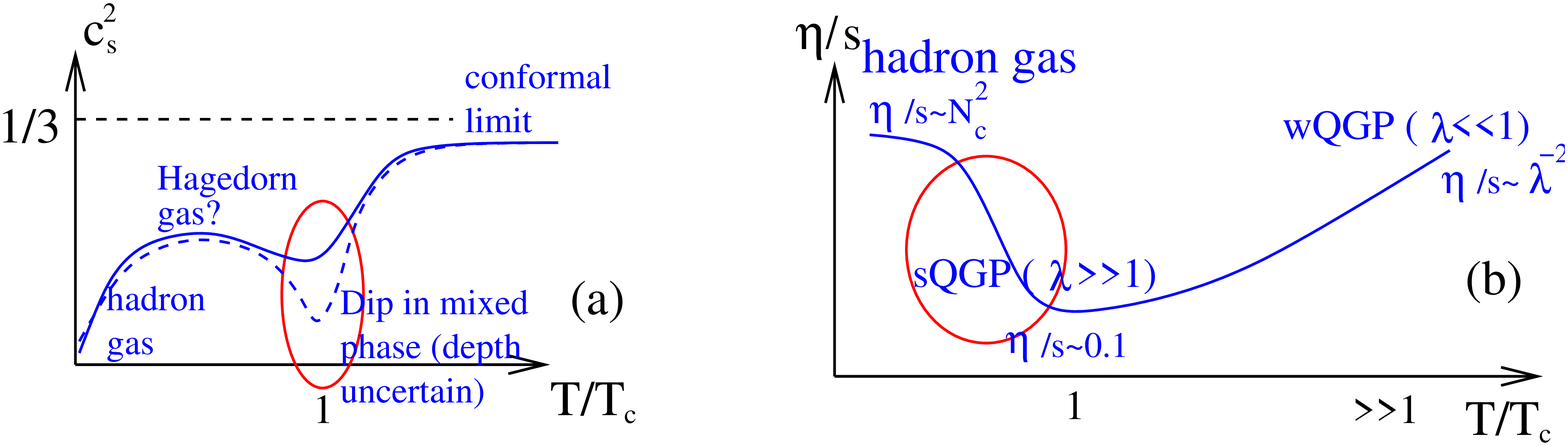}
\caption{\label{figT} (color online) Expectation of the change of the speed of sound (a), based on \cite{choj,jaki}), and the viscosity-to-entropy ratio $\eta/s$ with temperature ($N_c$ denotes the number of colors and $\lambda$ the  `t Hooft coupling constant). Red ellipses indicate the temperature regime where one could expect scaling violations to occur.}
\end{figure*}

Assuming the rapidity dependence of flow observables is ``encoded'' in the initial density (and associated intensive properties: $T,\eta/s$ etc.) rather than in the transverse size ($S \sim A^{2/3}$ at all $y$), and assuming subsequent evolution is local in $y$ (a natural assumption in the weakly interacting limit, less so in the strongly interacting limit, since  rarefaction waves can travel across rapidity \cite{mishustiny,bozek}), we should expect the scaling of $v_2$ to follow the scaling of $dN/dy$ {\em provided} the intensive properties are either invariant  with $\rho(\tau_{eq})$, or change monotonically with $\rho(\tau_{eq})$, throughout the rapidity range.   However, in a cross-over from a QGP to a hadron gas, intensive properties of the system should {\em not} change monotonically with $\rho(\tau_{eq})$.

In Fig. \ref{figT}, we summarize the expected changes:  The viscosity to entropy ratio $\eta/s$ is expected to jump from the relatively high value of the hadron gas ($\eta/s \sim N_c^2$ in a gas of mesons and glueballs, where $N_c$ is the number of colors \cite{thooft}) to the low value of strongly interacting QGP ($\eta/s \sim \order{0.1} N_c^0$ \cite{buchel}), and then slowly increase to the asymptotically free weakly coupled QGP ($\eta/s \sim \order{\lambda^{-2}} N_c^0$, where $\lambda$ is the `t Hooft coupling constant\cite{amy}).  In addition, the speed of sound is expected to have a dip in the cross-over region, whose depth is at the moment not well determined \cite{choj,jaki}.   

In a wide variety of models, elliptic flow $v_2$ depends on the eccentricity $\epsilon$, and should be approximately proportional to it. This essentially follows from Taylor-expanding the solution of whatever dynamical equation $v_2$ obeys in $\epsilon$, since
 $\epsilon$ is small and dimensionless, and since by symmetry with no $\epsilon$ there is no elliptic flow.   
\begin{figure}[t]
\epsfig{width=8cm,figure=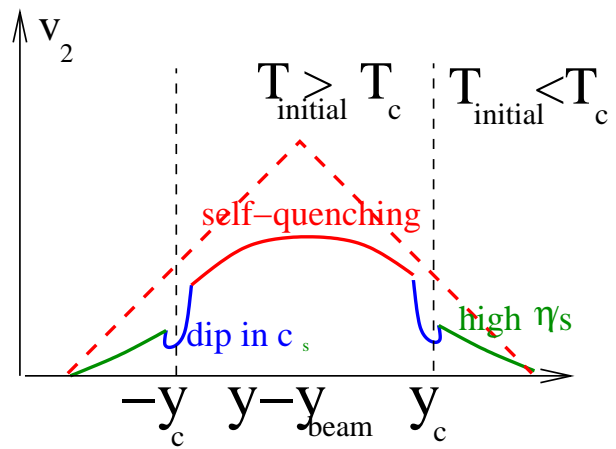}
\caption{\label{figv2} (color online) The $v_2$ dependence on rapidity given initial conditions reproducing limiting fragmentation, and subsequent hydrodynamic evolution. The superimposed dashed line shows the pre-equilibrium ($\tau=\tau_{dyn}$) partonic density from Fig. \ref{figsmear} (a) }
\end{figure}

We also know that $v_2/\epsilon$ decreases if viscosity is turned on, i.e. if $K$ increases.     Hence, it is quite natural that, as suggested in \cite{dumitru,scaling}
\begin{equation}
\label{dumeq}
\frac{v_2}{\epsilon}\sim \left. \frac{v_2}{\epsilon} \right|_{ideal} \left(1-\frac{K}{K_0} \right)  \simeq \left. \frac{v_2}{\epsilon} \right|_{ideal} \frac{K^{-1}}{K^{-1}+K_0^{-1}} 
\end{equation}
where $K$ is the Knudsen number and $K_0 \sim \order{1}$ is a parameter specific to the theory.
Furthermore, the transverse Knudsen number at a given mean free path $l_{mfp}$ is
\begin{equation}
\label{khydro}
K^{-1} \sim \l_{mfp}^{-1} \sqrt{S} \sim  \frac{ c_s}{ l_{mfp} S} \frac{dN}{dy}
\label{dumitru}
\end{equation}
This formula assumes just boost-invariant flow, as well as a a time-scale \cite{dumitru} $\tau_{v2}=\sqrt{S}/c_s $ for the building up of 
$v_2$, where $\sqrt{S}$ ($\sim N_{part}^{1/3}$, Not to be confused with the center of mass energy $\sqrt{s}$ or the entropy density $s$) is the initial transverse size of the system (which, as we argued earlier, is independent of rapidity).  
It should be noted that going beyond this rough approximation for $\tau_{v2}$ worsens scaling, since  \cite{v2orig,tauv2} $\tau_{v2}$ rises {\em and saturates} with increasing density due to the self-quenching of elliptic flow.

The derivation of Eqs. \ref{dumeq} and \ref{khydro} follows straight-forwardly \cite{dumitru} from  density formula \cite{bjorken}
\begin{equation}
\label{bjorkdens}
\rho \sim \frac{1}{S \tau_{eq}} \frac{dN}{dy}
\end{equation}
(where $S$ is the initial transverse surface and $\tau_{eq} \sim l_{mfp}$ )
and Taylor expanding

We further remember that $\left. \frac{v_2}{\epsilon} \right|_{ideal}$ depends on the equation of state, i.e. on the speed of sound.
By a leading order expansion argument, and remembering that the asymptotic expansion speed of a Godunov-type hydrodynamic shock wave $\sim c_s$ \cite{landau,taub,bouras}, it can be seen that $\left. 
\frac{v_2}{\epsilon} \right|_{ideal} \sim c_s$ (numerical simulations lend credence to this scaling, see \cite{dumitru}).   We also remember that $l_{mfp} \sim \frac{\eta}{T s}$. 

Putting everything together, and neglecting the difference between $y$ and the pseudo-rapidity $\eta$ (small in the fragmentation region away from mid-rapidity), we get that
\begin{equation}
\label{v2scalstrong}
\frac{v_2}{\epsilon} \sim c_s(\tau_{eq})\left(1 - \order{N_{part}^{-1/3}\mathrm{fm^{-1}}}\left[  \frac{c_s \eta}{T s} \right]_{\tau_{eq}} \right)
\end{equation}
we believe that when $T>T_c$ $\eta/s \ll 1$, $c_s \simeq 1/\sqrt{3}$, when $T\sim T_c$ $c_s \ll 1/\sqrt{3}$ and $\eta/s$ is at a minimum, and when $T<T_c$ $c_s$ goes back to a value not too different from $1/\sqrt{3}$ but $\eta/s$ increases to $\geq 1$.    (Fig. \ref{figT}).      

Additionally, close to mid-rapidity the plasma lifetime should $\gg \tau_{v2}$, so $v_2$ saturates and becomes independent of initial density (Eq. \ref{v2scalstrong} over-predicts $v_2/\epsilon$).  In the less dense region, however, the plasma lifetime $\leq \tau_{v2}$, so $v_2$ should be approximately proportional to the initial density (Eq. \ref{v2scalstrong} is a good approximation).

On the other hand, $N_{part}$ should be independent of rapidity and pseudorapidity  while the initial $T(\tau_{eq})$ should smoothly change as given by Eq. \ref{teq}.

We immediately see (Fig. \ref{figv2}) that the scaling seen in Fig. \ref{figeta} is {\em not} compatible with a modified BGK initial condition, or indeed any initial condition without an unphysically finely tuned correlation between the size of the system and intensive parameters \cite{scaling}.  In the supposedly long-lived ideal fluid mid-rapidity region, $v_2/\epsilon$ 
should be considerably flatter than $dN/dy$ due to the self-quenching of $v_2$ \footnote{Note that, as mentioned in the introduction, the existence and size of the mid-rapidity plateau of $v_2$ is currently controversial.  Even if it exists, however, it can not be much larger than one unit of rapidity}.
    At the critical rapidity where $T_{eq} \sim T_c$, $v_2/\epsilon$ should dip due to the dip in the speed of sound, and in the fragmentation regions where $T_0<T_{eq}$ $v_2/\epsilon$ should go down more rapidly than $dN/dy$ due to the rise in $\eta/s$.  The rapidity at which $v_2$ vanishes should in general be different from the rapidity at which $dN/dy$ does, due to $v_2$ additional dependence on $\eta/s$ and system lifetime.
These considerations are qualitative, and based on some simplifying assumptions ($c_s$ and $\eta/s$ do not change over the timescale $\tau_{v2}$, and stay constant significantly above and below $T_c$).  Because it is a scaling argument, however, more realistic dependence of $c_s$ and $\eta/s$ on temperature naturally leads to {\em more} scaling violation.

While no detailed hydrodynamical studies of energy scans with 3D 
hydrodynamics (ideal or viscous) have, so far, been performed, the 
studies of $v_2$ with pseudo-rapidity at the top RHIC energy 
\cite{v2rap1,v2rap2,v2rap3,v2rap4,v2rap5} confirm the insights of the present work. 

These simulations, while describing $v_2$ at mid-rapidity acceptably, 
have qualitative features absent from the data, such as fast changes in 
gradient of $v_2$ w.r.t. pseudorapidity (Fig. 10 of \cite{v2rap2}, Fig 3 
of \cite{v2rap3}). Some choices of initial conditions minimize these 
differences (\cite{v2rap1} Fig 3 and 4) but they never seem to 
disappear, as is natural due to the existence of intrinsic 
scales set by the phase transition and freeze-out conditions. 

These features are expected to become stronger if a temperature dependent viscosity (a la Fig. \ref{figT} (b)) is 
added to these models, and especially when non-equilibrium parameters, such as pre-thermalization flow (necessary for particle interferometry data) are included.
Experimental data, however, has no trace of these changes in gradient.

For one energy, this statement would only be as good as the non-negligible error bars of the experimental measurement.     The limiting fragmentation of $v_2$, however, would be spoiled had these variations of gradients been there.  Given that this limiting fragmentation {\em is} seen, spanning $\sqrt{s}=20-200$ GeV \cite{busza}, the evidence that $v_2$ drops uniformly with rapidity, with a gradient independent of $\sqrt{s}$, seems to be solid.   

An obvious way to make this scaling more natural is to assume that, rather than $\eta/s$ being constant, we have
\begin{equation}
\frac{\eta}{s} \sim \kappa \left( \frac{T_c^3}{s}\right)
\end{equation}
where $\kappa$ does {\em not} change across $T_c$.  Such behavior, however, contradicts the intuition from kinetic theory (where $\eta/s \sim T l_{mfp} \sim \order{\lambda^{-2}}\sim \order{T^0}$).
Furthermore, this solution is itself unnatural, since the equation of state experiences a cross-over and degrees of freedom change.  For example, for $SU(N_c)$ $\eta/s \sim N_c^0$ above $T_c$ and $\sim N_c^2$ below $T_c$.   Finally, this solution requires that the ``partonic triangle'' initial condition is also appropriate at lower (AGS) energies, where the initial temperature is according to the Bjorken formula $\leq T_c$ even at mid-rapidity. 

The observation of $v_2$ limiting fragmentation, therefore, poses a formidable challenge to the hydrodynamic model.

\section{Scaling of elliptic flow in the weakly interacting limit \label{secweak}}
The discussion in the previous section motivates us
to try to see if such a scaling could arise in {\em weakly} interacting systems, where collective effects play a smaller role and dynamics is determined {\em just} by the dilution and, rates of scattering and mean fields. 
The advantage of this is that the latter two could, in certain circumstances, be
less sensitive to whether the microscopic degrees of freedom at equilibration are partons or hadrons.  

In fact, a scaling of $v_2$ with density can be naturally derived by solving the Boltzmann equation with an ellipsoidal profile in initial transverse density, as was done in \cite{voloshin1}.  This gives rise naturally to a scaling relation of the form
\begin{equation}
\label{levy}
\frac{v_2}{\epsilon} \sim \frac{\ave{\sigma v}}{S} \frac{dN}{dy}
\end{equation} 
where $S$ is once again the transverse overlap region, and $\ave{\sigma v} \sim \l_{mfp}/\rho $ is the interaction cross-section between the degrees of freedom times their average relative velocity.   This scaling is surprisingly similar to that derived in Eq. \ref{dumitru}, except that $\sigma v$ is replaced by $\ave{\sigma} c_s $.  In a dilute system there is no reason, of course, to suppose $v$ is related to $c_s$ rather than to some initial intrinsic momentum.  For example, for massless particles $v=1$ while $c_s=1/\sqrt{3}$.

Together with the modified BGK scaling discussed previously, such an ansatz will naturally lead to a scaling of the type observed, {\em provided} that $\ave{\sigma v}$ does not vary with rapidity.   

This requirement might seem counter-intuitive:  At the critical density separating partonic and hadronic degrees of freedom, it is natural to expect  $\ave{\sigma v}$ to increase as the nearly massless color-charged partonic degrees of freedom  are liberated. This likely violation, in fact, was the original motivation for the derivation of \cite{voloshin1}.    

Considering, however, that the weakly interacting system will be far-away from equilibrium until hadronization (we note that the critical system size required for statistical hadronization is a topic of considerable controversy \cite{stat1,stat2}), and that breaking flux tubes emit partons rather than hadrons {\em throughout} the rapidity range of the system, it is not unreasonable to suppose  $\ave{\sigma v}$ maintains its partonic value even in the dilute large rapidity regions. 

A more serious problem is modeling the {\em absolute} value of $v_2$:  
as shown in \cite{molnar}, a model of the type described in  \cite{voloshin1} is not able to describe the {\em absolute} value of $v_2$ unless $\sigma$ was increased to the point where the mean Knudsen number is well below unity.   In such a regime, the approximation of ``one collision per particle per lifetime'', used to derive Eq. \ref{levy}, is clearly not appropriate and the ideal hydrodynamic limit emerges.

One can modify the derivation in \cite{voloshin1} with the introduction of a mean field.   This way, the magnitude of $v_2$ can be acceptably described (provided coalescence holds) even with a negligible $\sigma$ (the Vlasov equation limit) \cite{koch}, provided the lifetime of the system is long enough for the mean field description to be appropriate.  
If this is done, and the magnitude of the mean field is proportional to $dN/dy$ (a reasonable proposition), one can substitute  $\ave{\sigma v} \rightarrow \ave{M}$  in Eq. \ref{levy} ($\ave{M}$, the mean field per parton, being an independent constant) and increase $v_2$ far above the limits set by the $K\sim 1$ approximation.  The scaling naturalness inherent in Eq. \ref{levy} is maintained provided the system is, at formation, partonic (and described by Eq. \ref{bjorkdens}) throughout all rapidity.

  While such assumptions are not unreasonable since repulsive mean field potentials seem to be suggested by lattice QCD \cite{koch}, we are far away from assessing whether crucial aspects of this scenario (e.g. whether $\ave{M} \propto dN/dy$ at all energies, and whether it is really true that $v_2$ forms at times $\ll \tau_{eq}$) fit with QCD
 \cite{quasi1,quasi2,quasi3}.  We therefore limit ourselves with noting that such a model is the only one on the market where the strict proportionality of $v_2$ with $dN/dy$, and hence the limiting fragmentation of $v_2$, can be incorporated naturally.

\section{Discussion and conclusions \label{secexp}}
The unnaturalness of $v_2$ limiting fragmentation in the strongly interacting (hydrodynamic) limit, and the problems encountered to model $v_2$ in the weakly interacting (Vlasov equation) limit motivate us to look for {\em experimental} ways clarify the situation.
 
The LHC might well tell us very quickly if hydrodynamic descriptions of $v_2$ are appropriate, since the hydrodynamic prediction \cite{lhc1,lhc2,lhc3} differs from the extrapolation from limiting fragmentation \cite{busza,borglhc} by as much as $40 \%$.   The preliminary results suggesting that the multiplicity at mid-rapidity increases faster than $dN/dy \sim \ln \sqrt{s}$ \cite{cms,alice} suggest,if the ideas discussed in section \ref{subsec2} are correct, that limiting fragmentation at the LHC will break down to the same extent.  It will be very interesting to see if this is the case.

A promising endeavor for low energy measurements is to look
for limiting fragmentation in other soft observables.   In particular, it might be worthwhile to look for a similar limiting fragmentation in the average transverse momentum $\ave{p_T}$ \cite{brahmspt}, a break of which could signal the ``step'' \cite{vanhove,horn} in rapidity space. $\ave{p_T} \sim T + m_\pi v_T^2$, and $v_T$ should decrease significantly when the initial $T$ is around $T_c$ due to the dip in $c_s$ in the mixed phase regime.    Similarly, rapidity dependence of $R_{out}$ and $R_{side}$ HBT radii could be used to check whether the scaling seen in {\em global} $R_{out,side} \sim \left( \frac{dN}{dy} \right)^{1/3}$ radii \cite{lisahbt}  is also local in rapidity. Since accounting for $R_{out,side}$ at mid-rapidity with hydrodynamics requires adding flow created before equilibrium \cite{pratt}, limiting fragmentation of HBT radii will pose even greater challenges for hydrodynamics.  Such measurements will clarify whether the compressibility of the system does indeed change with rapidity, as its expected to in local thermal equilibrium where it mirrors the softness of the equation of state \cite{vanhove}, but not in a far-from equilibrium expansion driven by mean fields.
   
In this work, we have argued that the experimentally observed scaling of multiplicity with rapidity and $\sqrt{s}$ \cite{busza} follows from reasonable models of initial partonic dynamics.
Neither the free-streaming limit nor the ideal fluid limit are expected to break it.
The situation is however different with the equally simple scaling observed with $v_2$.   It is not clear how this scaling could arise within non-ideal hydrodynamics, even if its initial condition mirror closely the ones that reproduce the scaling observed in $dN/dy$.  The goodness of the scaling seen at RHIC, therefore, might force us to reconsider weakly interacting dynamics, where it arises more naturally.  Further measurements of rapidity dependence of other soft observables, within the context of an energy scan, might help clarify the situation further.   
G.T. acknowledges the financial support received from the Helmholtz International
Center for FAIR within the framework of the LOEWE program
(Landesoffensive zur Entwicklung Wissenschaftlich-\"Okonomischer
Exzellenz) launched by the State of Hesse.
We thank D. Rischke, M.Gyulassy, L. Ferroni, W. Florkowski, E. Santini, P. Steinberg, M. Bleicher, M.Gyulassy and J.Noronha for discussions and suggestions

\end{document}